\documentclass[fleqn,10pt]{wlscirep}
\usepackage[utf8]{inputenc}
\usepackage[T1]{fontenc}

\usepackage{graphicx}%
\usepackage{multirow}%
\usepackage{amsmath,amssymb,amsfonts}%
\usepackage{amsthm}%
\usepackage{mathrsfs}%
\usepackage[title]{appendix}%
\usepackage{xcolor}%
\usepackage{textcomp}%
\usepackage{manyfoot}%
\usepackage{booktabs}%
\usepackage{listings}%
\usepackage{anyfontsize}%
\usepackage[capitalise]{cleveref}%
\usepackage{makecell}
\usepackage{physics}%
\usepackage{lipsum} 
\usepackage{adjustbox}
\usepackage{subfig}
\usepackage{lineno}

\graphicspath{{./assets/}}

\title{Equitable non-contact infrared thermography after solar loading using deep learning}

\author[1,*]{Ellin Zhao}
\author[1]{Alexander Vilesov}
\author[1]{Pradyumna Chari}
\author[2]{Laleh Jalilian}
\author[1]{Achuta Kadambi}
\affil[1]{University of California, Los Angeles, Department of Electrical and Computer Engineering, Los Angeles, 90095, U.S.}
\affil[2]{University of California, Los Angeles, David Geffen School of Medicine, Los Angeles, 90095, U.S.}

\affil[*]{ellinz@ucla.edu}

\keywords{Health sensing, fever detection, facial thermal imaging}

\begin{abstract}

Widely deployed for fever detection, infrared thermometers (IRTs) enable rapid non-contact measurement of core body temperature but are inaccurate in unconstrained environments when skin temperature is transient. In this work, we present the first study on the effect of ``solar loading''--solar radiation-induced elevation of skin but not core temperature--on IRT performance. Solar loading causes poor specificity in IRT fever detection, and the standard procedure is to reacclimate subjects for up to 30 minutes before IRT measurement. In contrast, we propose a single-shot deep learning model that removes solar loading transients from thermal facial images, allowing accurate IRT operation in solar loaded conditions. Forehead skin temperature increases by 2.00°C after solar loading, and our deep learning model, SL-Net, reduces this error by 68\% to 0.64°C. We show that the solar loading effect depends on skin tone, introducing inequity in IRT performance, while SL-Net is unbiased. We open source a diverse dataset of 100 subjects with co-registered RGB-thermal images, and IRT and skin tone measurements. Our work shows that it is possible to use machine learning to correct complex thermal perturbations to enable robust and equitable human thermography.
\end{abstract}
\begin{document}

\flushbottom
\maketitle

\newcommand{\best}[1]{\textcolor{blue}{#1}}
\newcommand{\irt}[1]{IRT\textsubscript{#1}}


\begin{figure*}[th!]
    \centering
    \captionsetup{margin=1cm} 
    \includegraphics[width=0.8\linewidth]{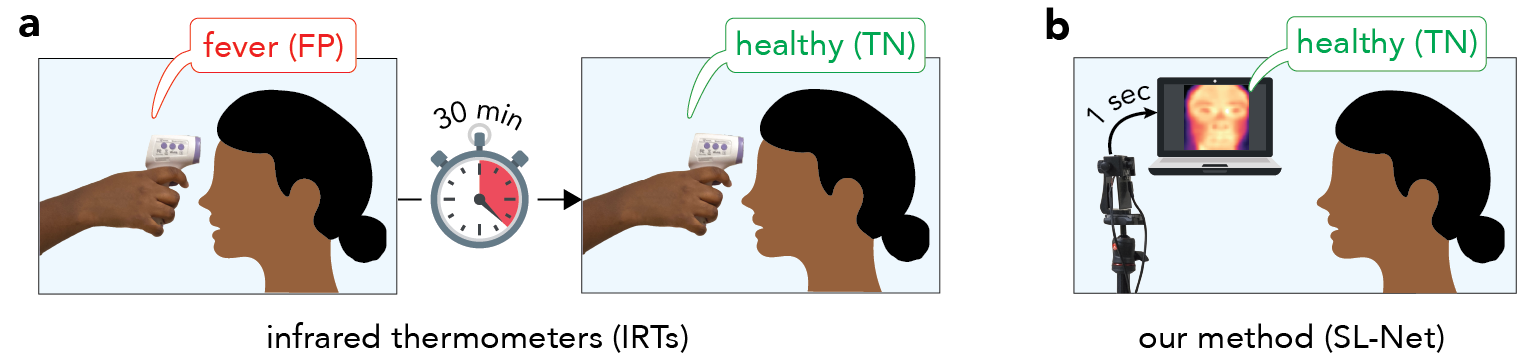} 
    \caption{
    Solar loading (SL) occurs when a subject is exposed to solar radiation, causing the skin to heat up while core body temperature remains the same.
    \textbf{a)} After SL, infrared thermometers (IRTs) report false positive (FP) fevers and the current solution to this problem is a 30 minute reacclimation period, after which IRTs report true negative (TN) febrile states.
    \textbf{b)} In contrast, our method (SL-Net) uses machine learning to remove SL effects from thermal data, and reports TN fevers in less than a second.
    }
    \label{fig:teaser}
\end{figure*}

\section{Introduction}\label{sec:intro}

Infrared thermometers (IRTs) enable rapid non-contact measurement of core body temperature, driving their popularity for mass fever screening during recent epidemics \cite{Sapra2024, Hussain2021}.
In a typical scenario of IRT use, a person enters a building after exposure to outdoor conditions, such as sunlight, and their access to the public space is determined by IRT fever screening. 
The IRT measurement, commonly performed at the forehead, is quick and contactless but has a major caveat—it is accurate only when a person is acclimated to room temperature conditions (draft-free, air temperature between 18-24°C, relative humidity between 20-75\%) \cite{IEC80601}. 
We call this acclimated state the \textit{baseline} temperature state. 
One common \textit{transient} (non-baseline) state that is the subject of this study is solar loading, in which solar radiation heats up the skin over a short period, causing transient skin temperature despite unchanged core body temperature.

IRTs are only calibrated to predict core body temperature when subjects are in the baseline state, but unfortunately devices are used on subjects in transient states.
To prevent IRT inaccuracies, manufacturers state that a subject must acclimate for up to 30 minutes prior to IRT measurement \cite{Daanen2020, adc, welch, sejoy}. During the acclimation period, skin and core body temperature transients dissipate, preventing inaccurate IRT readings, but the acclimation period is frequently ignored and reduces the utility of IRTs as rapid screening devices.

Despite the common occurrence of solar loading, there are no studies on how solar loading impacts IRTs or any methods for correcting solar loading errors beyond a reacclimation period. 
Our paper is the first to address these topics. We study IRT performance after solar loading and propose a machine learning method that corrects solar loading from a single facial thermal image, obviating the need for a long wait period. Our method corrects solar loading transients in less than a second—a vast improvement from a 30 minute acclimation period. 
The solar loading problem is summarized in \Cref{fig:teaser}.

Our solar loading correction model, SL-Net, is motivated by the insight that physiology (e.g. vasculature) dictates a standard baseline temperature appearance. Face temperature after solar loading also has a limited range of appearances due to physical constraints on sun angle and face geometry. 
When a person is solar loaded, their baseline skin temperature is obscured, but it is still present and can be extracted by a model that learns baseline (internal physiology), and solar loading (external heating) features.
Thermal physiological features, both hand-crafted and learned, have been successfully used for other applications such as face recognition, emotion recognition and intoxication detection \cite{Buddharaju2007, Lin2021, ClayWarner2014, Koukiou2009, Kristo2018}, and in this study we extend thermal features to health-based thermography. 

In our work, we fill a gap in research on IRT limitations \cite{Erenberk2013, Spindel2021, Qiu2023} and study how solar loading degrades IRT performance, and propose a single-shot method for solar loading correction.
We collect an extensive thermal dataset containing 100 subjects with measurements from multiple forehead IRTs, a thermal camera, a RGB camera, and a colormeter. 
Our dataset contains 144,000 co-registered RGB-thermal images and 900 thermometer measurements, which we use to show that solar loading (1) causes an error of 4°C in IRT temperature estimation, (2) is skin tone-dependent, and (3) can be corrected using machine learning, reducing the solar loading error by 68\%.
Our study on solar loading and the proposed correction model present a new paradigm for adapting infrared thermography to modern usages, enabling fair and robust remote health sensing that supports public health.

\section{Method} \label{sec:Method}

\begin{figure*}[p]
    \centering
    \includegraphics[width=1\linewidth]{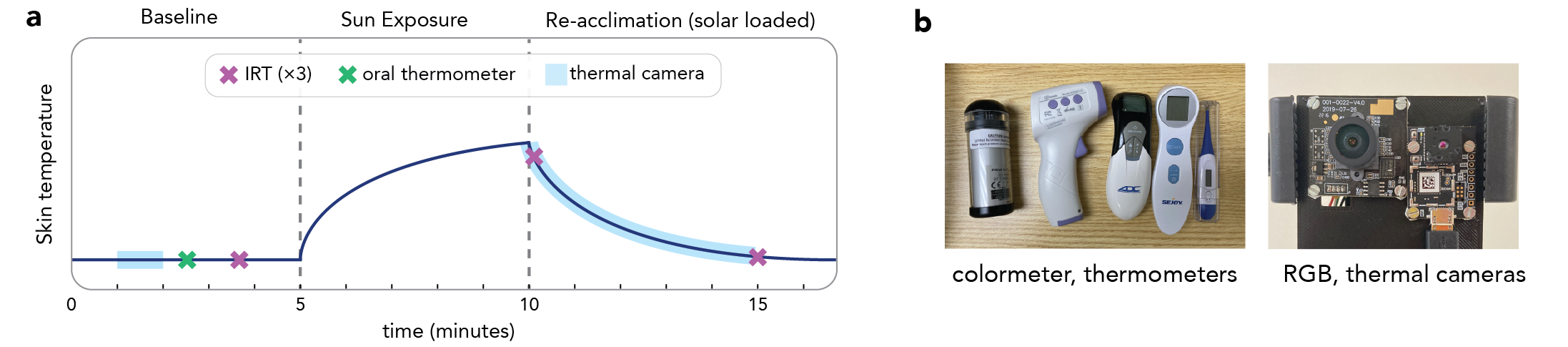}
    \caption{
    Overview of data collection.
    \textbf{a)} Timing of thermal measurements during the experiment. We collect IRT and thermal camera data during baseline and reacclimation stages. Oral temperature is taken once.
    \textbf{b)} Devices used in our data collection. We use a custom RGB-thermal camera setup, and multiple thermometers.
    }
    \label{fig:expt-timeline}
\end{figure*}

\begin{table*}[p]
\centering
\caption{
Summary of dataset values for $n=100$ subjects.
The mean values across experiment conditions, skin tone distribution, and thermometer measurements are shown with the measurement devices.
}
\label{tab:dataset-summary}
\begin{tabular}{llll}
    \toprule
    \textbf{Variable}                  & \textbf{Value}                 & \textbf{Total measurements} & \textbf{Measurement device} \\ \midrule
    \textbf{Experiment conditions}     &                                &                             &                             \\ 
    \quad Time of day                  & 12:42 PM $\pm$ 70 minutes      & 100                         & -                           \\
    \quad Ambient temperature          & 24.51 $\pm$ 1.94 °C            & 100                         & BTMeter 866A Anemometer     \\
    \quad Relative humidity            & 43.52 $\pm$ 12.38 \%           & 100                         & BTMeter 866A Anemometer     \\
    \quad Solar power                  & 1067.47 $\pm$ 107.16 Wm$^{-2}$ & 100                         & Tenmars TM-206 Solar Meter  \\ \midrule
    \textbf{Skin color}                &                                &                             &                             \\ 
    \quad Erythema                     & 17.59 $\pm$ 3.93 a.u.          & 100                         & Cortex DSM III Colormeter   \\
    \quad Melanin                      & 45.72 $\pm$ 11.52 a.u.         & 100                         & Cortex DSM III Colormeter   \\
    \quad Fitzpatrick Skin Type (I-VI) & 3.72 $\pm$ 1.21                & 100                         & Mean of 5 human annotators  \\ \midrule
    \textbf{Oral temperature}          & 36.53 $\pm$ 0.36 °C            & 100                         & Boncare MT-601A             \\
    \textbf{Infrared thermometer 1}    &                                &                             &                             \\
    \quad Baseline                     & 36.65 $\pm$ 0.22 °C            & 100                         & Welch Allyn 105801          \\
    \quad Solar loaded                 & 40.04 $\pm$ 1.44 °C            & 100                         & Welch Allyn 105801          \\
    \quad Cooled                       & 36.85 $\pm$ 0.29 °C            & 100                         & Welch Allyn 105801          \\
    \textbf{Infrared thermometer 2}    &                                &                             &                             \\
    \quad Baseline                     & 36.81 $\pm$ 0.17 °C            & 100                         & ADC Adtemp 429              \\
    \quad Solar loaded                 & 39.65 $\pm$ 1.41 °C            & 100                         & ADC Adtemp 429              \\
    \quad Cooled                       & 36.34 $\pm$ 5.65 °C            & 100                         & ADC Adtemp 429              \\
    \textbf{Infrared thermometer 3}    &                                &                             &                             \\
    \quad Baseline                     & 36.51 $\pm$ 0.57 °C            & 100                         & Joytech Sejoy DET-306       \\
    \quad Solar loaded                 & 40.65 $\pm$ 1.62 °C            & 100                         & Joytech Sejoy DET-306       \\
    \quad Cooled                       & 36.79 $\pm$ 0.59 °C            & 100                         & Joytech Sejoy DET-306       \\ \midrule
    \textbf{Thermal camera}            &                                &                             &                             \\
    \quad Baseline                     & - °C                           & 24,000 frames (160$\times$120)    & FLIR Lepton 3.5 LWIR        \\
    \quad Solar loaded                 & - °C                           & 120,000 frames (160$\times$120)   & FLIR Lepton 3.5 LWIR        \\
    \textbf{RGB camera}                &                                &                                  &                             \\
    \quad Baseline                     & -                              & 24,000 frames (1920$\times$1080)  & Arducam                     \\
    \quad Solar loaded                 & -                              & 120,000 frames (1920$\times$1080) & Arducam                     \\ \bottomrule
\end{tabular}
\end{table*}

\subsection{Dataset}

Data was collected from 100 subjects in baseline and solar loaded states.
All study procedures were approved by the UCLA Medical Institutional Review Board (IRB\#21-001473), and the study was conducted in accordance with the approved IRB and the Declaration of Helsinki.
Participants were mainly university students recruited using the REDCap system, and the dataset age range is 18-40 years. All participants signed an informed consent form prior to participation, and were made aware of their right to withdraw at any time. Informed consent was obtained to publish subject images.

\subsubsection{Experimental conditions}
Data is only collected under specific conditions to standardize solar loading for accurate bias analysis.
We collect data from 12:00-15:00 when it is sunny (solar power $> 900 \text{ Wm} \textsuperscript{-2}$), and there is minimal wind (wind speed $< 3 \text{ ms}\textsuperscript{-1}$).
We additionally record indoor air temperature and relative humidity using an anemometer. 
The measurement devices used and the average experiment conditions are listed in \Cref{tab:dataset-summary}.
Data collection is split into three sessions: baseline, sun exposure and reacclimation. The baseline and reacclimation sessions are completed indoors, and the sun exposure session is completed outdoors. The experiment timeline is shown in \Cref{fig:expt-timeline}a.

\subsubsection{Hardware setup}
Oral temperature is collected once during data collection.
We record skin tone at the forehead using the Fitzpatrick skin type (FST) determined by the average value from 5 annotators, and a colormeter, which is a handheld device that uses narrow-band reflectance spectrophotometry to report the erythema and melanin content in the skin.

Visible sweat is wiped from the face at all stages of the data collection.
A custom thermal-RGB camera setup with a sampling rate of 4 Hz is used to collect video data for 1 minute during the baseline, and for 5 minutes during reacclimation.
The camera is positioned 0.9 meters from the subject.
We collect measurements using 3 IRTs after the baseline, and at the beginning and end of reacclimation. 
IRTs are operated according to manufacturer instructions (held perpendicular to the forehead at a 2 cm measuring distance).
The IRTs used are shown in \Cref{fig:expt-timeline}b.

\subsubsection{Machine learning dataset}
In total, we have 900 IRT measurements and 144,000 frames of thermal data.
The RGB images are used to filter out thermal images where the face is not completely visible or not facing forward, and to determine facial regions-of-interest (ROIs) such as the forehead. Remaining thermal images are cropped to the face, and resized to size $32\times 32$.
Facial landmarks are extracted from the RGB facial images to obtain the forehead ROI, and reported forehead temperatures are the mean temperature over the ROI.
We only use the first 2 minutes of reacclimation data because the level of solar loading is ambiguous after a certain amount of cooling. 
Reacclimation (solar loaded) and baseline images are paired in preparation for model training.
For more details on data pre-processing, see the Supplementary Material.
We split our dataset into training and test sets, and the training set is sub-divided for 5-fold cross validation. The training set contains 82 subjects, with a total of 18,390 paired frames, and the test set contains 18 subjects, with 5,017 paired frames. Each subject appears exclusively in either the training or test set to prevent overfitting.
Random horizontal flips are applied to the images to augment the dataset.

\subsection{SL-Net: single-shot solar loading correction}

The SL-Net model consists of the reconstruction model, and an additional classifier to prevent over-correction of baseline inputs.
Our reconstruction model is a convolutional neural network (CNN) that takes in a thermal image and outputs estimates of the baseline and transient components.
The specific architecture is a residual U-Net architecture \cite{Liu2023} with 4 encoding and 4 decoding layers. 
The encoding layers reduce the input image into a feature map, and the decoder layers reconstruct the baseline and transient images from the features.
Each encoding layer consists of 2 convolution modules (2D convolution, instance normalization, and ReLU activation), and a skip connection (2D convolution and batch normalization).
Each decoding layer follows a similar structure with the addition of bilinear upsampling before the convolution modules.
The classifier is input the feature map from the final encoder layer of the Residual U-Net, and classifies whether the input is transient or baseline. If the input is a baseline image, the reconstructed image is ignored. The classifier is a simple fully-connected network with 1 hidden linear layer of size 256, and ReLU activations following all layers. 

The residual U-Net is trained using a combination of losses. To evaluate the accuracy of the temperature estimate, we calculate the mean absolute error (MAE) on the model output and the ground truth baseline image.
The paired solar loaded and baseline images are not perfectly aligned, which incurs a double penalty from the MAE loss, so we include Learned Perceptual Image Patch Similarity (LPIPS) loss in our training \cite{Zhang2018_LPIPS} to ensure the model focuses on similar thermal appearance and not alignment errors. The MAE and LPIPS losses are weighted equally.

The classifier is optimized using binary cross-entropy loss.
Both the classifier and reconstruction models are trained using the AdamW optimizer \cite{AdamW} with a learning rate of 0.005. 
Data processing code is written in Python and the machine learning code is created using the PyTorch library in Python.

\subsection{SL-FCN: ROI regression}
We develop a simple spatial model, SL-FCN, that corrects solar loading using ROI temperatures across the face. 
We use the forehead, left and right inner canthi, and left and right cheek. The forehead and inner canthi have higher blood flow, and tend to be closer to core body temperature, while the cheeks show greater influence from the environment \cite{Foster2021}. 
The ROIs are localized using facial landmarks detected from the RGB images.
The ROI model is a fully-connected network (FCN) with one hidden layer of size 16. 
1D batch normalization and ReLU activations are applied after each linear layer. This model is optimized with the same MAE losses used for the Residual U-Net.

\subsection{Temporal extrapolation}
After solar loading, skin temperature reacclimates to indoor conditions and decays to a steady state, where the steady state is the desired baseline temperature.
In general, cooling (and heating) can be described by an exponential function.
Given a time series of temperature at the forehead, $T(t)$, cooling is described by:
\begin{equation}
    T(t) = A e^{-Bt} + T_0,
    \label{eq:temporal}
\end{equation}
where $A$ is the maximum solar loading increase, $B$ is the rate of cooling which depends on environmental temperature and other factors, and $ T_0$ is the baseline.
We extract forehead ROI over time from the thermal images, and fit this data to \Cref{eq:temporal}.
The variables are constrained to physically possible values: $32\leq  T_0<40.5$°C, $0 \leq A < 5$°C and $0 \leq B < 1$.
Curve fitting is performed using a trust-region reflective algorithm \cite{scipy}.

\section{Results}\label{sec:results}
\newcolumntype{C}[1]{>{\centering\let\newline\\\arraybackslash\hspace{0pt}}m{#1}}

\begin{table}[t!]
    \centering
    \caption{
    Mean absolute error (°C) between infrared thermometer (IRT) and oral thermometer values during the baseline, and after 0 and 5 minutes of reacclimation after SL.}
    \begin{tabular}{lccc}
    \toprule
    \textbf{Device} & \textbf{Baseline (°C)} & \textbf{Reacclimation at 0 min (°C)} & \textbf{Reacclimation at 5 min (°C)} \\ \midrule
    \irt{1} \cite{welch} & 0.28 $\pm$ 0.26  & 3.52 $\pm$ 1.45  & 0.40 $\pm$ 0.37  \\
    \irt{2} \cite{adc}   & 0.34 $\pm$ 0.30  & 3.15 $\pm$ 1.41  & 0.44 $\pm$ 0.39  \\
    \irt{3} \cite{sejoy} & 0.38 $\pm$ 0.49  & 4.14 $\pm$ 1.64  & 0.50 $\pm$ 0.48  \\ \bottomrule
    \end{tabular}
    \label{tab:irt_results}
\end{table}

In this study we separate the problems of baseline \textit{skin} temperature and \textit{body} temperature estimation, and while we study IRT body temperature errors, our proposed solution operates on skin temperature. For this reason, we do not directly compare our methods to IRTs because they return different variables.

\subsection{IRT inaccuracies after solar loading}
IRTs are known to perform poorly in non-baseline environments \cite{Spindel2021, Erenberk2013, Dzien2020, Ogawa2022, Ravi2022, Tay2015}, but no studies have evaluated the accuracy of IRTs after sunlight heats up the skin at IRT measurement sites.
Our study on $n=100$ subjects provides the first results on this effect.
This study defines solar loading as sun exposure that results in elevated skin temperature and unchanged core body temperature, and we use a controlled 5 minute sun exposure period to emulate this. Prolonged sun exposure is out of scope because core body temperature may change (i.e. hyperthermia).
\label{sec:sl-net-scope}
Oral temperatures taken before ($36.55\pm 0.23$°C) and after 5 minutes of solar loading ($36.62\pm 0.27$°C) confirm no significant change in oral body temperature as measured by a paired t-test ($t(16)=-0.85, p=0.41$).

We calculate the mean absolute error (MAE) between three different IRTs and a ground truth oral thermometer.
The laboratory accuracy for all 3 IRTs follow the ISO 80601-2-59 \cite{IEC80601} and ASTM E1965-98 \cite{astm} standard of $\pm 0.3$°C accuracy. In practice, the operational accuracy for our application is reported to be $\pm 2- 5\%$ of the measured temperature \cite{welch, adc, sejoy}.
The oral thermometer reports an accuracy of $\pm 0.2$°C \cite{boncare_therm}.

IRT errors are within the stated range, and the errors both after solar loading and after 5 minutes of reacclimation are higher than the baseline error, as shown in \Cref{tab:irt_results}.
During the baseline, \irt{1} (MAE of 0.28°C), \irt{2} (0.34°C) and \irt{3} (0.38°C) have errors within their stated ranges.
After solar loading, the errors for all three IRTs increase nearly tenfold to a range of 3.15-4.14°C despite oral body temperature remaining the same.
After 5 minutes of reacclimation, the IRT error decreases to 0.40-1.02°C, which is still higher than the baseline error.
Five minutes of reacclimation is shorter than the suggested 30 minutes, but these results are still meaningful because they establish that a shorter but practical reacclimation period is not sufficient for mitigating IRT errors.

\begin{figure*}[t!]
    \centering
    \captionsetup{margin=1cm} 
    \includegraphics[width=\linewidth]{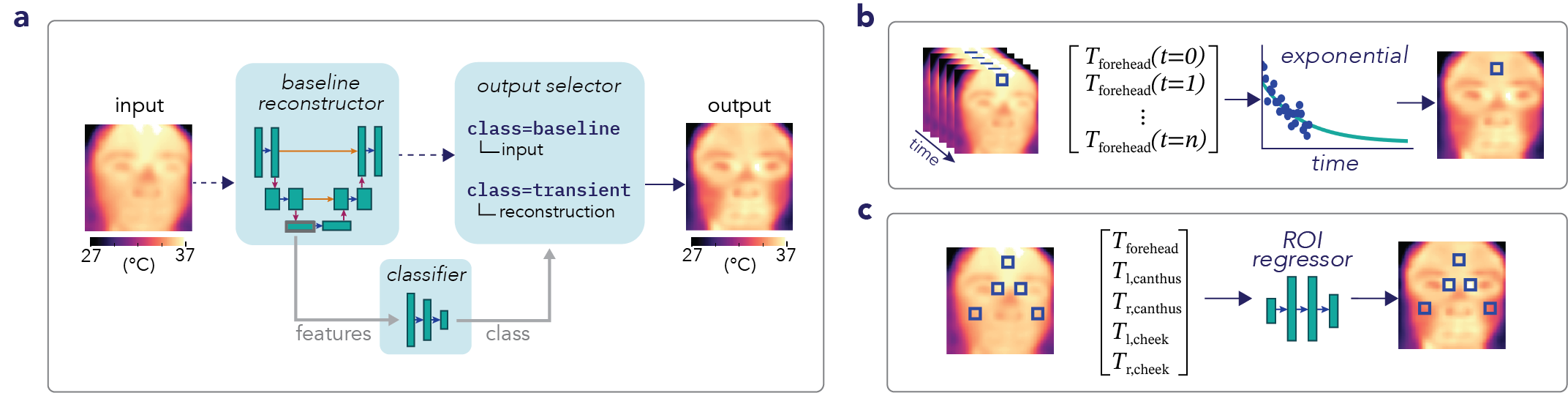} 
    \caption{
    We propose 3 methods of baseline estimation to circumvent the 30 minute reacclimation period for IRTs. 
    \textbf{a)} Our best method, SL-Net, uses a reconstruction network and a classifier on a single frame of thermal data.
    \textbf{b)} The temporal extrapolation method fits 5 minutes of forehead temperature to a skin cooling model.
    \textbf{c)} SL-FCN, a fully-connected network, uses 5 region-of-interest (ROI) temperatures (forehead, left and right canthi, left and right cheeks) extracted from a single thermal image.
    }
    \label{fig:methods}
\end{figure*}

\begin{table*}[t!]
    \centering
    \caption{Our SL-Net model significantly reduces the error incurred from solar loading. Results are calculated on multiple measurements from $n=18$ subjects using mean absolute error (MAE), mean squared error (MSE), mean absolute percentage error (MAPE).}%
    \begin{tabular}{lcccc}
    \toprule
    \textbf{Method} & \textbf{MAE (°C)} & \textbf{RMSE (°C)} & \textbf{MAPE (\%)} \\ \midrule
    No correction                 & 2.00 $\pm$ 0.94     & 2.21          & 5.94 $\pm$ 2.84 \\
    Ours (temporal extrapolation) & 1.38 $\pm$ 0.71     & 1.55          & 4.09 $\pm$ 2.11 \\
    Ours (SL-FCN)                 & 0.70 $\pm$ 0.41     & 0.81          & 2.09 $\pm$ 1.25 \\
    \textbf{Ours (SL-Net)}        & \textbf{0.64$\pm$ 0.42} & \textbf{0.76} & \textbf{1.93 $\pm$ 1.21} \\ \bottomrule
    \end{tabular}
    \label{tab:solar_forehead_results}
\end{table*}

\subsection{The SL-Net model}
We develop a deep neural network, SL-Net, to estimate baseline skin temperature (equivalently, to remove solar loading temperature transients) from a thermal image. The thermal camera we use is inexpensive (\$200) and has an accuracy of $\pm 5$°C \cite{Lepton}.
Instead of requiring a 30 minute reacclimation period \cite{Daanen2020}, our proposed SL-Net produces accurate sub-second baseline skin temperature estimation by removing transients from a thermal image. The SL-Net predicted baseline skin temperature can be passed to a downstream body temperature estimator, but skin-to-body temperature conversion is not addressed in this work.
SL-Net has two parts: a reconstruction model that estimates the baseline temperature from the input thermal image, and a classifier that determines whether the input image is transient and requires correction, as shown in \cref{fig:methods}a.

\subsection{SL-Net forehead temperature estimation} \label{sec:sl-net-performance}
SL-Net accurately predicts baseline forehead temperature on our dataset of solar loaded images.
Our dataset consists of paired baseline and solar loaded thermal images from 100 subjects. The dataset is split subject-wise into train (82 subjects) and test (18 subjects) sets, and we report results from the test set subjects that are unseen during training.
To evaluate SL-Net, we compute the error between the estimated and ground truth baseline forehead temperature, and compare these results against the error incurred from applying no correction.

We choose our final model based on cross-validation results across 5 dataset splits. The cross-validation MAE for solar loaded images is $0.696 \pm 0.097$.
For our test dataset, SL-Net model significantly reduces the solar loading temperature error across multiple metrics as shown in \Cref{tab:solar_forehead_results}. Without correction, the mean SL temperature increase (after 2 minutes of reacclimation) is 2.00°C, and the SL-Net model achieves a MAE of 0.64°C, corresponding to a 68\% reduction in error from the no correction case. Across all other error metrics (root mean squared error, mean absolute percentage error), SL-Net reduces the no correction error by more than 65\%.

\begin{figure*}[th!]
    \centering
    \captionsetup{margin=1cm} 
    \includegraphics[width=0.75\textwidth]{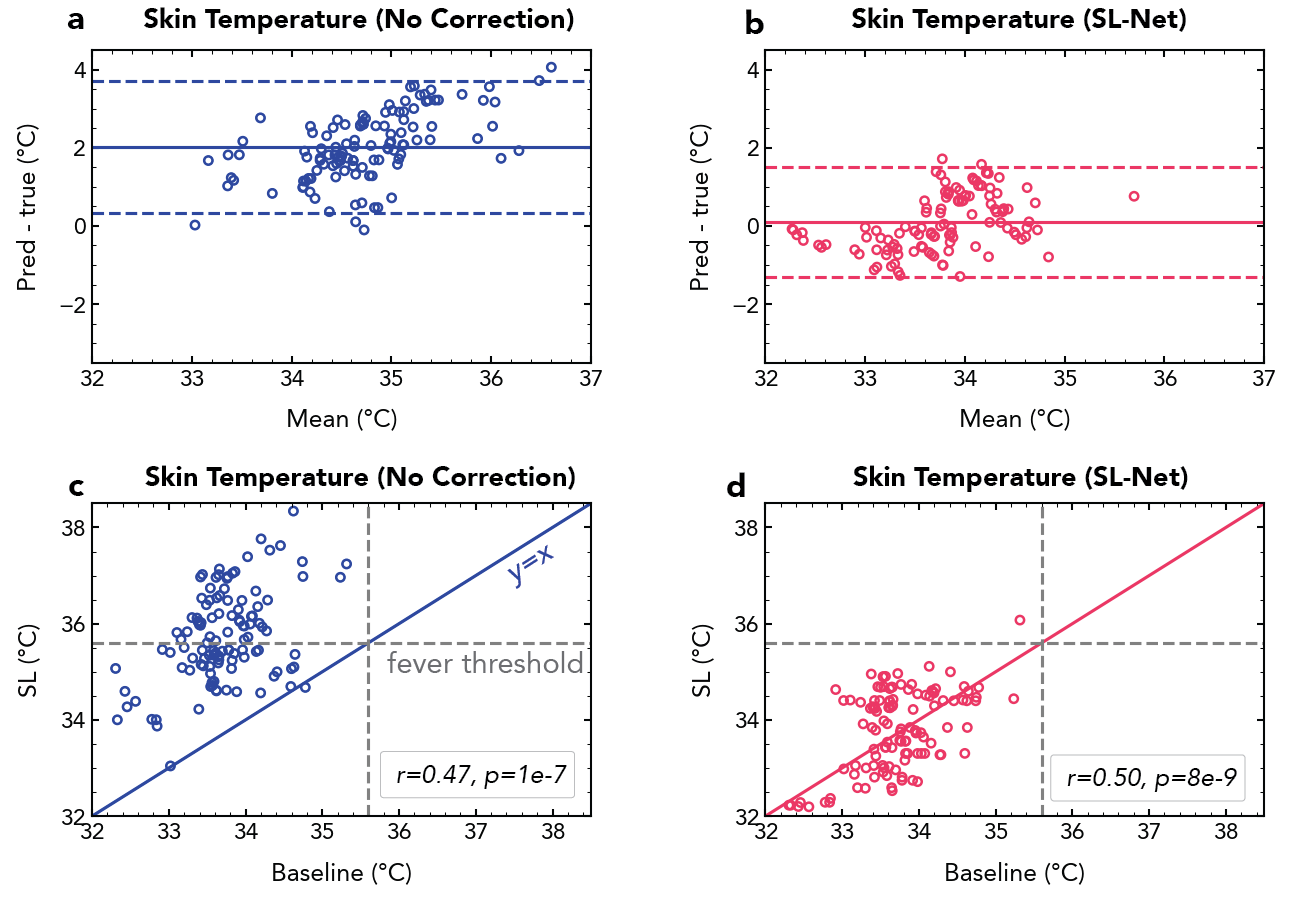}
    \caption{
    SL-Net corrected temperatures achieve better correlation with the ground truth and reduce false positive fever detection compared to uncorrected SL temperatures.
    Bland-Altman plots for solar loaded and baseline temperatures with \textbf{a)} no correction and \textbf{b)} SL-Net correction indicate that SL-Net achieve lower errors. The solid line is the mean between the predicted and ground truth, and dashed lines represent $\pm$1.96 SD.
    Correlation plots for solar loaded and baseline temperatures with \textbf{c)} no correction and \textbf{d)} SL-Net correction, where the dashed gray lines shows fever threshold of 35.6°C \cite{Ng2005} and the solid line shows the line of agreement.
    Pearson correlation coefficient ($r$) is reported.
    }
    \label{fig:bland-altman}
\end{figure*}

\subsubsection{Comparison to other baseline estimation methods}
We compare SL-Net to two additional methods of baseline estimation in \Cref{tab:solar_forehead_results}.
To the best of our knowledge, no works have addressed solar loading correction, so we develop additional methods inspired by recently proposed IRT methodologies \cite{Qiu2023, Shajkofci2022, Silawan2018}.
Our first method is \textit{temporal extrapolation} (\cref{fig:methods}b), which fits temporal forehead data to a decaying exponential and uses the steady state term as the baseline estimate.
Our second method is \textit{SL-FCN} (\cref{fig:methods}c), in which we extract temperatures from 5 regions-of-interest (ROIs) (forehead, left and right inner canthi, left and right cheek), and train a fully-connected network (FCN) to estimate the baseline ROI temperatures. 

The methods that use spatial input data (SL-FCN, SL-Net) outperform temporal extrapolation, and SL-Net achieves the lowest errors out of all evaluated methods.
The temporal method uses 5 minutes of forehead temperature data, and performs the worst with a MAE of 1.38°C.
In contrast, the spatial methods, SL-FCN and SL-Net, use only a single thermal frame of data from inference and achieve lower errors (0.70, 0.64°C respectively).

\subsubsection{Agreement plots}
We further analyze the agreement between the solar loaded temperatures against the ground truth baseline temperature using agreement plots.
The Bland-Altman plots for the uncorrected and SL-Net corrected forehead temperatures are shown in \Cref{fig:bland-altman}a and b respectively.
Without correction, the solar loading error is high and has a large spread (2.00$\pm$0.86°C) because the SL transient decreases exponentially during reacclimation. 
With SL-Net correction, the mean difference between the prediction and ground truth is 0.099°C, and the errors have a smaller spread ($\pm$0.70°C).

We show the correlation between the baseline estimates (uncorrected, SL-Net corrected) and ground truth in \Cref{fig:bland-altman}c and d.
Despite the large difference between solar loaded and baseline temperatures, the two variables remain moderately correlated (Pearson $r=0.47, p\approx 0$) as shown in \Cref{fig:bland-altman}c. 
SL-Net correction slightly improves the correlation ($r=0.50, p\approx 0$) as shown in \Cref{fig:bland-altman}d.
The skin temperature fever threshold of 35.6°C \cite{Ng2005} is shown in the correlation plots, and reveals that uncorrected forehead temperature can easily be mistaken for a fever. Without correction, nearly half of the datapoints would be erroneously classified as febrile (top-left quadrant in \cref{fig:bland-altman}c), resulting in poor fever detection specificity. Using SL-Net, only one datapoint would be erroneously classified as febrile (\cref{fig:bland-altman}d), and all other points are correctly classified as non-febrile.
To quantify agreement between the 3 IRTs we report the intraclass correlation coefficient (ICC). We evaluate a fixed number of IRTs, so we report the two-way mixed, single measures, consistency ICC metric \cite{Koo2016}. For the baseline, solar loading and acclimated states, the ICC values are 0.56, 0.54, and 0.29 respectively. The agreement between IRTs is lowest after 5 minutes of acclimation, while the ICC for the baseline and solar loaded cases are similar.

\begin{figure}[t]%
    \centering
    \captionsetup{margin=0.1cm} 
    \subfloat[Body temperature from oral and infrared thermometers (IRTs) (°C). \label{tab:irt-temp-states}]{
    \begin{tabular}{lcccc}
        \toprule
        \textbf{Device} & \textbf{Baseline} & \textbf{Solar loaded} & \textbf{Hot air} & \textbf{Exercise} \\ \midrule
        Oral                 & 36.83    & 37.00    & 37.00   & 37.33   \\ 
        \irt{1} \cite{welch} & 37.00    & 40.33    & 37.45   & 36.61   \\
        \irt{2} \cite{adc}   & 37.00    & 41.28    & 37.61   & 36.89   \\
        \irt{3} \cite{sejoy} & 37.33    & 41.89    & 37.50   & 36.56   \\ \bottomrule
        \end{tabular}
    }
    \qquad
    \subfloat[Skin temperature and SL-Net outputs (°C). \label{fig:recons}]{
    \includegraphics[width=0.5\linewidth]{
        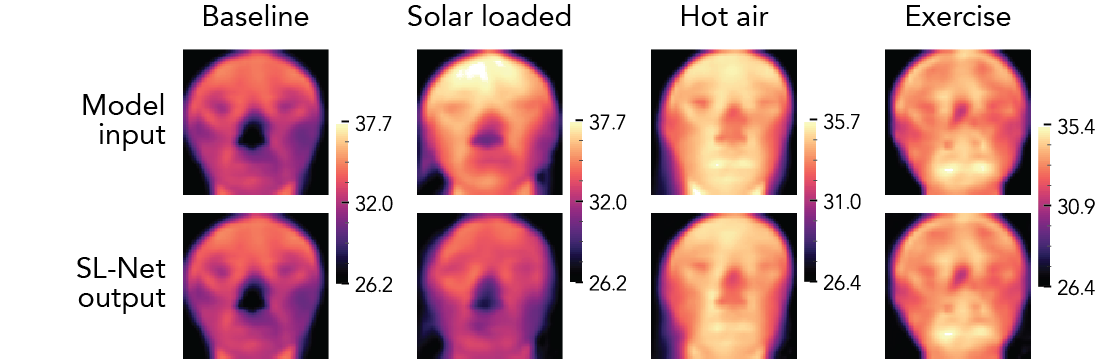}
    }
    \caption{Temperature for one subject for baseline, solar loaded (5 min sun exposure), hot air exposure (26.7°C for 30 minutes) and exercise (indoor aerobics) states.}
\end{figure}

\subsection{SL-Net generalization to unseen temperature states}
SL-Net is trained only on baseline and solar loaded data, and we verify that SL-Net does not over-correct unseen skin and oral body temperature states through a longitudinal study.
We collect data from one subject in 4 temperature states: baseline, after 5 minutes of solar loading, after 30 minutes of hot air (26.7°C) exposure, and after 20 minutes of indoor aerobic exercise. 
Body and skin temperature for these states are shown in \cref{tab:irt-temp-states} and \cref{fig:recons} respectively.

Relative to the baseline, oral temperature increases after solar loading (+0.17°C), hot air exposure (+0.17°C), and exercise (+0.5°C). 
We note that solar loading does not lead to oral body temperature increase in general (over multiple subjects) as discussed in \Cref{sec:sl-net-scope}. Previous studies confirm that core body temperature increases after hot air exposure \cite{Haldane1905} and exercise \cite{Lim2008}. 
IRTs overestimate after both solar loading and hot air exposure, but underestimate after exercise.
SL-Net only corrects inputs that meet our definition of solar loading (skin temperature increase, no oral body temperature change) so only the solar loaded input is corrected. 
The corrected solar loaded image (\cref{fig:recons}) matches the appearance of the baseline, except for some detail from the supratrochlear artery.
After hot air exposure and exercise, oral body temperature increases, so correcting the input to return the baseline would not be accurate because the baseline reflects a cooler body temperature.
Elevated body temperature should correspond to an increase in skin temperature (as is the case for fevers), but we see that perspiration-induced evaporative cooling results in local regions of colder skin temperature after exercise.
For this subject, SL-Net is able to handle exercise and hot air inputs because it is not overfitted to solar loading data.
Our work focuses on solar loading, so the analysis on unseen temperature states is a supplemental result. Future works can evaluate more temperature states on a larger study population.

\begin{figure*}[t]
    \centering
    \captionsetup{margin=1cm} 
    \includegraphics[width=0.88\linewidth]{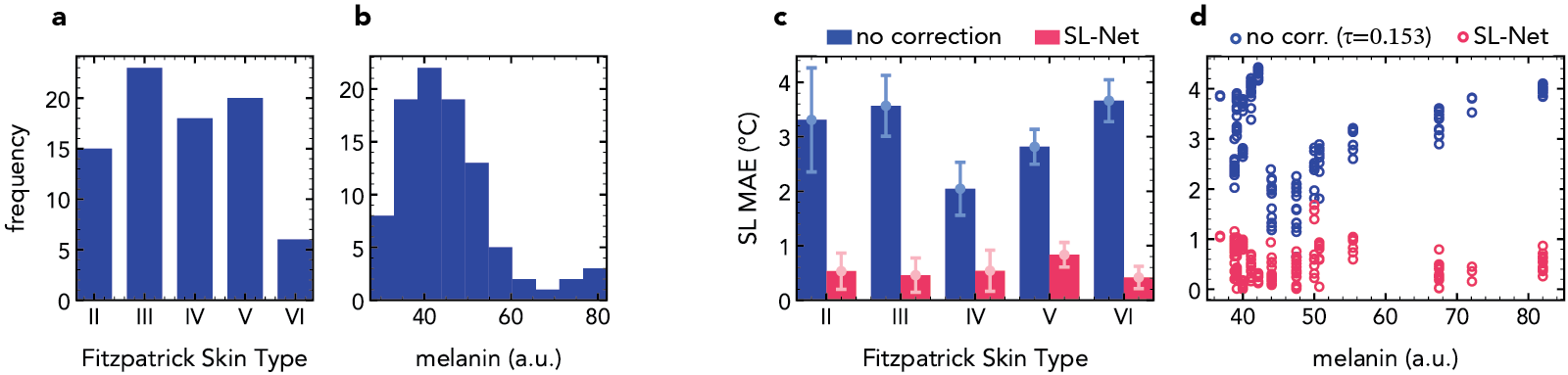}
    \caption{
    Our dataset is diverse in skin tone, which is measured using \textbf{a)} Fitzpatrick Skin Type (FST) and \textbf{b)} melanin levels from a colormeter.
    There is no correlation between solar loading errors and FST in \textbf{c)}, while melanin levels are correlated with SL errors in \textbf{d)} (Kendall-$\tau=0.153$). SL-Net outputs are not correlated with skin tone.}
    \label{fig:bias-skin-tone}
\end{figure*}

\subsection{Mitigating skin tone bias after solar loading}
We show that solar loading is a skin tone-dependent effect and that SL-Net removes this bias.
Some studies on IRT bias use coarse, subjective measures of skin tone (e.g. light, medium, dark \cite{Khan2021}) while others use ethnicity or race as a proxy for skin tone \cite{Strasse2022}, leading to conflicting results with studies focusing on racial biases and not skin tone biases \cite{Bhavani2022}.
Our study differs from prior work because we concretely define skin tone as melanin concentration determined by a colormeter, and this objective measure of skin tone allows us to accurately analyze bias.
We record Fitzpatrick skin type (FST) (Type I-VI), and show its limitations \cite{Ware2020} in our bias analysis.
We collected data for all FST categories except FST I, and the median FST of the subjects is Type IV (IQR: III-V) (\cref{fig:bias-skin-tone}a).
The range of melanin values is 27.58-82.02 with a median value of 43.03 (IQR: 37.48-50.59) (\cref{fig:bias-skin-tone}b).
FST and melanin distributions are similar except for the right tail where melanin has more granularity than FST. 
We define skin tone bias as a positive correlation between level of skin pigmentation and solar loading error. 
Correlation is measured by the Kendall rank correlation coefficient ($\tau$).

We evaluate bias on uncorrected and SL-Net corrected solar loading temperatures. 
No bias is found when comparing FST and no correction error ($\tau=-0.06, p=0.26$), and when comparing FST and SL-Net correction ($\tau=-0.02, p=0.71$) as shown in \Cref{fig:bias-skin-tone}c.
Evaluating using melanin reveals a different result. 
\Cref{fig:bias-skin-tone}d shows a positive correlation between melanin and no correction error, albeit a low correlation ($\tau=0.15, p=0.002$). 
There is no correlation between melanin and SL-Net errors ($\tau=-0.07, p=0.14$).
For IRTs, the reference measurement is the oral thermometer value.
Two IRTs exhibit skin tone bias: \irt{1} error is correlated with skin tone ($\tau = 0.23, p=0.0009$) and \irt{3} error is correlated with skin tone ($\tau = 0.24, p=0.0008$), but \irt{2} shows no bias ($\tau = 0.17, p=0.01$).

\section{Discussion} \label{sec:discussion}

In this work, we study how solar loading—the heating of skin through solar radiation—impacts infrared thermometer performance, and we propose a machine learning model that recovers baseline temperature after solar loading.
Our approach corrects solar loaded skin temperature in under a second while IRTs remain inaccurate for at least 5 minutes after cooling.
Our key findings are that 
(1) commercial infrared thermometers are highly inaccurate after solar loading, (2) it is possible to correct solar loading using machine learning, and (3) solar loading exhibits a skin tone bias that our model corrects.
Our work adds to recent research on limitations of IRTs and unlike previous studies, we present a solution to the limitations.

In our study, we compare IRTs to ground truth oral thermometer measurements in various temperature states. 
While oral thermometers are not the gold standard device for contact-based core body temperature measurement, they are more accurate than IRTs \cite{Jensen2000} and are the most suitable method of contact-based body temperature estimation for our experimental setup.
We evaluate IRT performance after solar loading when subjects are reacclimating to indoor conditions. During this state, it is not recommended to operate IRTs and device performance is expected to be poor. 
No single IRT performs the best; \irt{1} is best during baseline (MAE 0.28°C) and after 5 minutes of reacclimation (MAE 0.40°C), and \irt{2} is best at the start of reacclimation (MAE 3.15°C).
For a single subject, we evaluated the same IRTs for additional temperature states: hot air (26.7°C) acclimation and exercise. The IRTs overestimated after hot air acclimation by 0.52°C on average.
Our results for hot air and exercise states are consistent with prior works \cite{Chen2022, Fernandes2014}.
The effects of hot air and exercise are not the main subject of this study, but these preliminary results on a single subject are provided to establish that IRTs may be inaccurate in these states and that SL-Net does not overfit to this subject's solar loading data.
Our work uses a controlled solar loading session of 5 minutes, and we define solar loading to be skin heating that does not affect the core temperature. Our results can potentially be extrapolated to longer solar loading durations, but longer heating durations may result in hyperthermia. Future work will study the limits of solar loading durations to which our method still applies.

We find that the temporal method for solar loading correction is the least effective, despite the effect being time-varying by definition. The temporal extrapolation performs the worst despite taking a time series of 1,200 datapoints (5 minutes) of forehead cooling data as input.
The temporal extrapolation model is a decaying exponential that represents skin reacclimation and is a simplified version of a skin cooling model derived from the bioheat equation \cite{JSteketee1979}.
In theory, we only need 3 datapoints to fit to an exponential, but empirically this was not feasible and even with 5 minutes of data, the extrapolation was inaccurate.
One particular issue preventing correction in the same timeframe as the spatial methods (sub-second) is measurement error: the thermal camera's sensitivity is too low to resolve the temporal gradients seen during reacclimation.
Specifically, our camera has an noise equivalent temperature difference (NETD) of 0.05°C \cite{Lepton}. 
If we assume someone's temperature is raised by 2°C (based on the mean increase in \cref{tab:solar_forehead_results}), and that they cool completely in 5 minutes, then forehead temperature is $T(t) = 2e^{-0.025t}+ T_0$ and $\frac{dT}{dt}=-0.05 e^{-0.025t}$, where $T_0$ is the baseline (steady state) temperature. 
Using only 1 second of measurements to correct solar loading, we need $|\frac{dT}{dt}| \geq \texttt{NETD} = 0.05$, which is only true at $t=0$. It usually takes longer than 5 minutes to completely reacclimate, so theoretically it is not possible to correct solar loading in a second using noisy temporal data, and empirically not possible to correct solar loading using minutes worth of data likely due to model mismatch.

We show that spatial thermal features can be used to detect and remove temperature transients.
Our key insight is that temperature states have canonical appearances due to physiology and face geometry. Baseline skin temperature is largely determined by facial vasculature, while solar loading depends on face geometry because it is an external effect. 
The forward model for both baseline and solar loaded temperatures are described by the bioheat equation \cite{Charny1992}, but it is difficult to correct solar loading using this forward model because it is a system of partial differential equations that depend on unknown spatially-varying skin parameters.
See the Supplementary Material for more details on the bioheat equation.
Instead of inverting for baseline temperature from a forward model, we use a machine learning model to automatically detect physiological patterns hidden by the solar loading transient.
Prior works using thermal data have leveraged physiology without using the bioheat equation \cite{Buddharaju2007, Lin2021, ClayWarner2014, Koukiou2009, Kristo2018}, but this has not been done for transient correction in infrared thermography.
Introducing new variables (i.e. heart rate) to SL-Net can provide an opportunity to incorporate skin temperature physics from the bioheat model. 
Vital signs can be estimated using remote photoplethysmography (RPPG) on RGB images, which we record but do not utilize in our model.

We develop and evaluate two spatial models for correcting solar loading: a fully-connected network (FCN) operating on face ROIs, and a convolutional neural network (CNN) that processes an image of the face.
The FCN is input a hand-crafted feature of 5 temperature ROIs (forehead, left and right inner canthi, left and right cheek), and achieves an error that is 65\% lower than using no correction, and 49\% lower than the temporal method, despite using 0.6\% of the number of inputs that the temporal model uses.
SL-Net takes in the full face image, which is processed using a convolutional neural network (CNN), and further reduces the error to 0.67°C, and takes in 1,024 datapoints (32$\times$32 image), which is less than the temporal input size. 
The CNN is the most complex model, and is well suited for processing skin temperature, which is governed by complex dynamics. 
The improvement in error from the temporal method to spatial methods indicate that spatial data encodes meaningful physiological information for transient correction. 
Our spatial methods, which require only a single frame of data, are practical for public health screening due to their rapid inference times. However, single frame data cannot uniquely characterize temporal information, and our model relies on approximate spatial features that convey temporal information. There is potential for similar thermal patterns to map to different baseline states due to inter-individual variability in melanin, vasodilation and other variables. We examine how SL-Net errors vary across one variable, melanin, and find that SL-Net performance is not biased towards any specific category of skin color. Future work can examine other variables with potential confounding effects, such as blood flow, and examine how additional input variables can further improve model performance.

\label{sec:discussion-bias}
In non-contact health sensing applications, it is critical to consider how skin appearance can impact device accuracy.
Biases arise when devices are not built to handle a diverse range of skin tones (e.g. pulse oximeters \cite{Kadambi2021, Sjoding2020}). 
In the case of solar loading, skin tone bias arises from the physical process of melanin absorbing solar radiation, so we conduct bias analysis with melanin measurements, which allows us to separate skin tone bias from racial physiological biases. 
We show that melanin and solar loading increase have a low positive correlation (\cref{fig:bias-skin-tone}d), a correlation that is not discernible when using FST to measure skin tone (\cref{fig:bias-skin-tone}c).
For accurate bias analysis, we control for factors that affect solar loading (solar power, wind, etc), but we cannot control for thermoregulation.
Sweating is a means of thermoregulation that noticeably cools the skin (as seen in the exercise example in \cref{fig:recons}).
Melanin protects sweat glands from UV rays, so it has been proposed that people with darker skin have better thermoregulation \cite{Lambert2008}, which weakens the correlation between solar loading and skin tone.
Our SL-Net model removes correlation between skin tone and solar loading error, but has no notion of skin tone; simply addressing solar loading is enough to resolve the bias.

We have shown that our SL-Net model is capable of removing solar loading effects and preliminary results on one subject show that it does not over-correct unseen temperature states. 
We do not consider febrile states in this study—our dataset only contains afebrile subjects because the study setup made it difficult to collect data from febrile subjects before they received anti-pyretic treatment in a hospital.
Future work can extend our dataset by collecting data from febrile subjects with core temperatures above 37°C, and the addition of febrile data will ensure better sensitivity in SL-Net fever detection.
SL-Net has a quick inference time and improves specificity in fever detection; combined with febrile data, SL-Net can be a powerful tool for rapid health screening in highly-populated public areas, such as airports and hospitals.

Our method is easy to integrate due to the inexpensive hardware setup and software-based thermal analysis.
Our hardware setup uses an inexpensive thermal camera (\$200) and a RGB camera. The RGB camera is used to extract the forehead ROI, but recent work has been done in thermal ROI detection \cite{Chu2019}, making the RGB camera optional.
Screening systems that are already equipped with thermal cameras only need to integrate the software for SL-Net inference.
Our current dataset only uses data from one thermal camera, and future work can modify SL-Net to generalize to multiple thermal camera models in a data-driven manner by expanding the training dataset to contain data from multiple camera, or in a physics-drive manner by removing camera-specifica thermal noise \cite{Saragadam2021}.

In addition to SL-Net, this work open-sources an extensive dataset with 100 subjects of diverse skin tones. 
Our dataset may be useful for studying the robustness \cite{Hermosilla2012} of thermal methods (facial recognition, emotion recognition) after solar loading.
Our work shows promising results in improving infrared thermography under transient conditions, and our dataset enables more work to be done in this area.


\bibliography{main}

\section*{Author contributions}
A.K. conceived of the presented idea. L.J. and A.K. supervised the project.  L.J. supervised on medical aspects of the project. A.V., P.C., and L.J. designed the study. A.V. and P.C. developed the initial theory and implementation of the code.  E.Z., A.V. and P.C. performed the experiments.  E.Z. and A.V. developed the machine learning code. E.Z. took the lead in writing the manuscript. All authors provided critical feedback and helped shape the research, analysis and manuscript.

\section*{Funding}
This material is based upon work supported by the National Science Foundation
Graduate Research Fellowship Program under Grant No. DGE-2034835. Any
opinions, findings, and conclusions or recommendations expressed in this material
are those of the author(s) and do not necessarily reflect the views of the National
Science Foundation.

\section*{Data and code availability}
The dataset presented in this work is collected as part of an IRB study and data will be released according to those guidelines.
The following data are available via Harvard Dataverse (https://doi.org/10.7910/DVN/YG5KC2): (1) anonymized colormeter, oral thermometer and IRT measurements after solar loading, and (2) sample thermal images for the subject of the longitudinal study across temperature states is also provided, but without RGB images.
Access to the full dataset (with RGB images) for research purposes will be granted.
Please contact the corresponding author for further information.
The RGB-thermal camera calibration and operation code is available on GitHub (https://doi.org/10.5281/zenodo.17982320). 
The dataset preparation, model training and analysis code is also available on GitHub (https://doi.org/10.5281/zenodo.17982394).

\end{document}